\documentclass{INTERSPEECH2023}

\interspeechcameraready 

\usepackage{amssymb}
\usepackage{amsmath}
\usepackage{multirow}
\usepackage{multicol}
\usepackage{caption} 
\usepackage{algorithm}
\usepackage{algorithmicx}
\usepackage{dirtytalk}
\usepackage{algpseudocode}
\usepackage{graphicx} 
\usepackage{amsfonts}
\usepackage{caption}
\usepackage{subcaption}
\usepackage{censor}  

\def\L{{\cal L}}
\def\D{{\cal D}}
\def\B{{\cal B}}
\def\T{{\cal T}}
\def\M{{\cal M}}

\title{Rehearsal-Free Online Continual Learning for Automatic Speech Recognition}
\name{Steven Vander Eeckt, Hugo Van hamme}
\address{
  KU Leuven \\ 
     Department Electrical Engineering ESAT-PSI, Leuven, Belgium}
\email{\{steven.vandereeckt, hugo.vanhamme\}@esat.kuleuven.be}

\begin{document}

\maketitle
 
\begin{abstract}
Fine-tuning an Automatic Speech Recognition (ASR) model to new domains results in degradation on original domains, referred to as Catastrophic Forgetting (CF). Continual Learning (CL) attempts to train ASR models without suffering from CF. While in ASR, offline CL is usually considered, online CL is a more realistic but also more challenging scenario where the model, unlike in offline CL, does not know when a task boundary occurs. Rehearsal-based methods, which store previously seen utterances in a memory, are often considered for online CL, in ASR and other research domains. However, recent research has shown that weight averaging is an effective method for offline CL in ASR. Based on this result, we propose, in this paper, a rehearsal-free method applicable for online CL. Our method outperforms all baselines, including rehearsal-based methods, in two experiments. Our method is a next step towards general CL for ASR, which should enable CL in all scenarios with few if any constraints. 
\end{abstract}

\section{Introduction}
Catastrophic Forgetting (CF) \cite{catastrophicforgetting} occurs when Automatic Speech Recognition (ASR) models are extended to new domains (e.g. accents, languages, speakers, topics, etc.) which differ from the original domain the models were trained on. It means that by learning the new domains, the models' performance of the original domain degrades. This severely limits the possibility to build very powerful, diverse and inclusive ASR models, performing well on all dialects, accents, speakers, topics, etc. because the ASR models cannot be properly extended. Learning a new dialect, accent or speaker will result in the model forgetting old dialects, accents or speakers.  

Continual Learning (CL) attempts to find strategies to train models without suffering from CF. Within ASR, CL has recently been gaining attention \cite{lifelongasr, eeckt2021continual, eeckt_adapters, sustek22_interspeech, updating_only, disentangle, ogem, weight_averaging}. However, with the exception of \cite{ogem}, the focus of the above research is on offline CL rather than the more challenging online CL.

In offline CL, the model, trained on an initial task, is extended to new tasks. Tasks are represented by training and validation sets which the model has access to until it has learned the given tasks. Moreover, for the model, it is clear when one tasks ends and another starts.  
In online CL, on other hand, the model receives a stream of batches which it has to process. Once a batch has been learned, access is lost. Moreover, the model does not know whether two consecutive batches belong to the same task or not, i.e. it does not know when a task boundary occurs. Clearly, online CL is a more realistic and generally applicable though also more challenging scenario than offline CL. 

To the best of our knowledge, \cite{ogem} is the only work considering online CL for ASR. In \cite{ogem}, a well-known CL method from computer vision, Gradient Episodic Memory (GEM) \cite{gem}, is applied to online CL for ASR and referred to as O-GEM (Online GEM). O-GEM, like GEM, is a rehearsal-based method, which means that it attempts to overcome CF by storing previously seen utterances in a small memory. Utterances in this small memory are then later used during training of new batches to prevent forgetting.  
In computer vision, online CL has received much attention, with most of the proposed methods also being rehearsal-based methods \cite{er, aljundi_ocl, dark_er, cls_er, synergy}, since the gap between rehearsal-based and rehearsal-free methods, which do not use a memory, remains, in particular for online CL, large. 

The same could be said about offline CL in ASR \cite{lifelongasr, eeckt2021continual}. However, recently, \cite{weight_averaging} found weight averaging (i.e. computing the average of the model before and after being adapted to a new task) to be very effective. Without using a memory, their simple method outperformed rehearsal-based methods; which is significant because storing utterances from previous tasks is not always allowed nor desired. Nevertheless, their method is not applicable to online CL. Based on the simple but very effective method from \cite{weight_averaging} and inspired by the methods from \cite{cls_er, synergy} for computer vision, we propose an online CL method that is rehearsal-free and uses weight averaging. In two experiments, our method outperforms the rehearsal-based method from \cite{ogem} as well as the rehearsal-based methods applicable to online CL from \cite{eeckt2021continual}. We believe that this paper is an important next step towards general CL \cite{dark_er} for ASR, which must enable CL in all scenarios and with few if any constraints, since our method achieves the best performance without requiring a memory and without requiring to know the task boundaries.    

\section{Model}
We consider an encoder-decoder end-to-end ASR model with parameters $\theta \in \mathbb{R}^N$, taking as input speech frames $X$ of size $L_F \times d_i$ with $L_F$ the number of frames. The output tokens of the model are $C$ word pieces. Given ground truth $y$ of $L_W$ outputs tokens, the model's loss consists of a cross-entropy (CE) loss $\L_\text{dec}$, computed on the output of the decoder, and a CTC loss $\L_\text{ctc}$, computed on the output of the encoder (with $0 \leq c \leq 1$):
\begin{equation}
    \L_\text{ce}(X, y; \theta) = (1 - c) \L_\text{dec}(X, y; \theta) + c \L_\text{ctc}(X, y; \theta)
    \label{eq:ce_loss}
\end{equation}

\section{Online Continual Learning}
\begin{table*}
    \centering
    \caption{Results after learning the stream of batches from Fig. \ref{fig:subfig:s1}. All WERs (expressed in percentages) are evaluated on the final model. $\dagger$ indicates that the method's hyper-parameters were optimized on the test experiment. Best AWER result is in bold.   }
    \begin{tabular}{l c c c c c c c c r}
    & & & \multicolumn{6}{c}{\textit{WER per task}}  \\
    \cmidrule(lr){4-9} 
    Model & $M$ & $(\tau, \lambda, \tau_2)$ & $\T_0$--US & $\T_1$--ENG & $\T_2$--AUS & $\T_3$--IND & $\T_4$--SCO & $\T_5$--IRE & \textbf{AWER}  \\
    \toprule
    \multicolumn{3}{l}{Initial model $\theta_0$} & 17.3 & 13.9 & 15.4 & 21.4 & 15.2 & 11.0 & 15.72 \\
    \multicolumn{3}{l}{FT} & 19.4 & 15.0 & 15.2 & 26.1 & 15.5 & 11.5 & 17.12   \\
    \midrule 
    UOE & & & 18.3 & 12.6 & 13.1 & 21.6 & 14.2 & 11.0 & 15.15 \\
    EWC$\dagger$ & & & 18.3 & 13.4 & 14.1 & 22.4 & 14.7 & 11.1 & 15.66 \\
    ER$\dagger$ & $0.5k$ &  & 18.1 & 13.2 & 13.9 & 22.1 & 14.7 & 11.1 & 15.50 \\
    ER$\dagger$ & $2.0k$ &  & 17.9 & 12.9 & 13.6 & 21.5 & 14.1 & 10.8 & 15.13 \\
    O-GEM & $2.0k$ & & 19.0 & 14.1 & 14.4 & 25.2 & 14.9 & 11.3 & 16.48  \\
    \midrule 
    AOS & & $(1, 0.1, 1)$ & 17.1 & 12.7 & 14.0 & 21.4 & 14.3 & 10.7 & 15.03 \\
    AOS$\dagger$ & & $(2, 0.1, 1)$ & 17.3 & 12.5 & 13.6 & 21.5 & 14.2 & 10.7 & \textbf{14.96} \\
    \bottomrule
    \end{tabular}
    \label{tab:seq1}
\end{table*}

The objective of continual learning is to learn new tasks without forgetting old ones. Often, it is assumed that tasks boundaries are known and that the data of the tasks remains available until the task has been learned by the model; this is called offline CL. However, for online CL, this is not the case.  Access to each sample or batch is lost once it has been seen by the model and the task boundaries are unknown. 

Formally, a model $\theta_0$ has been trained on an initial task $\T_0$ with data $\D_0$, containing $D_0$ samples. Next, the model receives a (non-i.i.d.) stream of batches $\B_i$ ($i > 0$) of size $B$, which it processes batch after batch. When learning $\B_i$, access to $\B_{i-1}$ is lost, except possibly through a memory $\M_{i-1}$ of fixed size $M$. Batches belong to a certain task $\T_{t_i}$, but this is not known by the model. The model should learn all batches well while retaining the knowledge from old batches and tasks.

\section{Method}

Our method, which we call AOS (\textbf{A}veraging for \textbf{O}nline CL of A\textbf{S}R), consists of two parts: averaging (Sec. \ref{subsec:oa}) and regularization (Sec. \ref{subsec:reg}). An overview is given in Algorithm \ref{alg:sqnsampling}.  
\subsection{Online Averaging}
\label{subsec:oa}
Inspired by the effectiveness of weight averaging for CL in ASR \cite{weight_averaging} and the methods of \cite{cls_er, synergy}, we consider what we refer to as 'online averaging' for our online CL method. In \cite{weight_averaging}, the model is first adapted to a new task, after which the average was computed between the model before and after adaptation. The weight of the adapted model is $\eta=1/t$ with $t$ the number of seen tasks. However, in \cite{weight_averaging}, task boundaries are assumed to be known, which is not the case here. Therefore, we consider 'online averaging', i.e. we average the 'old' and 'adapted' model after each batch. In other words, if $\theta_{i}$ is the 'final' model at batch $\B_i$, and $\tilde{\theta}_{i+1}$ is the model adapted to a batch $\B_{i+1}$, then the final model after batch $\B_{i+1}$ is:
\begin{equation}
    \theta_{i+1}=(1-\eta_{i+1})\theta_{i} + \eta_{i+1} \tilde{\theta}_{i+1}
    \label{eq:avg}
\end{equation}
Two questions remain here. 

\textit{The adapted model.} What is $\tilde{\theta}_{i+1}$?  If $\tilde{\theta}_{i+1}$ is $\theta_{i}$ trained on a new batch $\B_{i+1}$, then $\theta_{i+1}$, instead of through Eq. \ref{eq:avg}, could be obtained by learning the new batch with a $\eta_{i+1}$ times smaller learning rate than $\tilde{\theta}_{i+1}$, i.e. our method would just be fine-tuning on the new batch with a $\eta_{i+1}$ times smaller learning rate. We do not expect this to work well. Therefore, we keep two models: one, $\theta_{i+1}$, which we call the final model; and another, $\tilde{\theta}_{i+1}$, the adapted model, which starts from $\theta_0$ but is then adapted to the new batches and thus from this point on deviates from $\theta_i$. It is very likely that the adapted model suffers from forgetting; at inference time, only the final model should thus be considered. The adapted model is only to aid the final model by transferring new information to it. This resembles the methods from \cite{cls_er, synergy}, which focus on computer vision and were inspired by the Complementary Learning Systems theory \cite{cls}. 

\textit{Weighted average.} What is the value of $\eta_{i+1}$? Most similar to $\eta=1/t$ from \cite{weight_averaging} would be $\eta_{i+1} = B / (D_{i} + B)$, where $D_{i}$ is the amount of data the model has seen so far and $B$ is the current batch size.  However, we make a number of improvements:
\begin{enumerate}
    \item The model receives as input utterances consisting of a number of frames, $L_F$. To allow longer utterances to have an higher impact, we consider $F$, the number of frames in a batch, rather than the batch size $B$. $F_i$ is then the number of frames seen after processing batch $i$, and $F_0$ is the total number of frames of initial task $\T_0$. 
    \item For the decoder of the model, the length of an utterance is related to its number of output tokens $L_W$. Therefore, for the decoder, we consider $W$, the number of output tokens in a batch, rather than the number of frames $F$; similar to $F_i$ and $F_0$, $W_i$ and $W_0$ are then the total number of output tokens after processing batch $i$ and the number of output tokens in initial task $\T_0$, respectively. This means that the encoder and decoder have separate values $\eta_{\text{enc},i+1}$ and $\eta_{\text{dec}, i+1}$ to compute the average in Eq. \ref{eq:avg} after processing batch $i+1$.
    \item If the original task was a very large one, i.e. if $F_0$ ($W_0$) is very large, then $\eta_{\text{enc}, i+1}$ ($\eta_{\text{dec}, i+1}$) will be very small and it will take a long time for the encoder (decoder) of the model to learn new information. Therefore, we introduce $\tau\geq1$ to increase the plasticity of the model:
    \begin{equation}
        \eta_{\text{enc}, i+1} = \frac{\tau \cdot F}{F_{i} + \tau \cdot F}
        \label{eq:eta}
    \end{equation}
    and the same for $\eta_{\text{dec}, i+1}$, where we then consider $W$ and $W_i$. 
    \item For monolingual experiments, the encoder is more prone to forgetting than the decoder \cite{eeckt_adapters}. On the other hand, \cite{updating_only} shows that freezing the decoder might overcome forgetting. Therefore, we consider $\tau_2$ for $\eta_{\text{dec}, i+1}$: 
    \begin{equation}
        \eta_{\text{dec}, i+1} = \frac{\tau_2 \cdot W}{W_{i} + \tau_2 \cdot W}
        \label{eq:eta_dec}
    \end{equation}
    With $(\tau, \tau_2)$, the decoder might be updated more conservatively ($\tau_2 \leq \tau)$ or progressively ($\tau_2 \geq \tau$) than the encoder. 
\end{enumerate}
In summary, the final model $\theta_{i+1}$ is obtained after averaging (Eq. \ref{eq:avg}) with the adapted model $\tilde{\theta}_{i+1}$, which is trained on the stream of batches, using the weight $\eta_{\text{enc}, i+1}$ from Eq. \ref{eq:eta} for the encoder and $\eta_{\text{dec}, i+1}$ from Eq. \ref{eq:eta_dec} for the decoder.

\begin{algorithm}
\caption{AOS (\textbf{A}veraging for \textbf{O}nline CL of A\textbf{S}R)}
\label{alg:sqnsampling}
\begin{algorithmic}[1]
\State Given: initial model $\theta_0$ trained on data $\D_0$ with $D_0$ utterances, $F_0$ frames and $W_0$ output tokens. 
\State Choose: $\alpha$, $c$ (for model); $\tau$, $\lambda$, $\tau_2$ (for AOS)
\State Set: $i \gets 0$, $\tilde{\theta}_0 \gets \theta_0$
\State \textit{\# online continual learning:}
\For{each batch $(X, y) \in \B_{i+1}$}
\State \textit{\# compute the loss on new batch for adapted model}
\State $\L \gets \L(X, y; \tilde{\theta}_i)$ \Comment{See Eq. \ref{eq:loss}}
\State \textit{\# update adapted model with SGD}
\State $\tilde{\theta}_{i+1} \gets \tilde{\theta}_i - \alpha \nabla \L$
\State \textit{\# $F$: number of frames in $X$}
\State $\eta_{\text{enc},i+1} \gets \tau F / (F_i + \tau F)$ \Comment{See Eq. \ref{eq:eta}}
\State \textit{\# $W$: number of output tokens in $y$}
\State $\eta_{\text{dec},i+1} \gets \tau_2 W / (W_i + \tau_2 W)$ \Comment{See Eq. \ref{eq:eta_dec}}
\State \textit{\# use $\eta_{\text{dec},i+1}$ for decoder parameters, else use $\eta_{\text{enc},i+1}$}
\State $\eta_{i+1} \gets \eta_{\text{dec},i+1}\text{ if decoder layer else }\eta_{\text{enc},i+1}$
\State \textit{\# update the final model}
\State $\theta_{i+1} \gets (1 - \eta_{i+1}) \theta_{i} + \eta_{i+1} \tilde{\theta}_{i+1}$  \Comment{See Eq. \ref{eq:avg}}
\State \textit{\# update seen frames $F_i$ and seen outputs $W_i$}
\State $(F_{i+1},\text{  } W_{i+1}) \gets (F_i + F,\text{  }  W_i + W)$
\State $i \gets i + 1$
\EndFor
\end{algorithmic}
\end{algorithm}

\subsection{Regularization}
\label{subsec:reg}
We apply regularization to the adapted model to improve the performance of the final model. We consider knowledge distillation (KD) \cite{knowledge_distillation} as in Learning without Forgetting (LWF) \cite{lwf}, a popular CL method that uses the data of the current batch to distill knowledge from the old to the current model. With $\hat{s}^i_{\text{ctc}, kc}$ and ${s}^i_{\text{ctc}, kc}$ the CTC softmax output of the $c$th word piece at the $k$th frame of, respectively, the final and adapted model after batch $i$, the KD loss for the CTC output becomes:
\begin{equation}
        \L_\text{ctc, kd} (X; \theta) = \sum_{k=1}^{L_F} \sum_{c=1}^C \hat{s}^i_{\text{ctc}, kc} \log {s}^i_{\text{ctc}, kc}
        \label{eq:kd_loss_ctc}
\end{equation}
The KD loss for the decoder, $\L_\text{dec, kd}(X, y; \theta)$, is computed similarly, by replacing, in Eq. \ref{eq:kd_loss_ctc}, the CTC softmax outputs by the decoder softmax outputs, with the outer sum then summing over all $L_W$ outputs. Since the decoder is autoregressive, its KD loss depends on the ground truth $y$. Overall, the KD loss becomes:
\begin{equation}
    \L_\text{kd}(X, y; \theta) = (1-c)\L_\text{dec, kd}(X, y; \theta) + c\L_\text{ctc, kd} (X; \theta)
\end{equation}
With $\lambda$ the regularization weight, the above KD regularization loss is then added to the model's loss from Eq. \ref{eq:ce_loss} as follows:
\begin{equation}
    \L(X, y; \theta) = (1 - \lambda)\L_\text{ce}(X, y; \theta) + \lambda \L_\text{kd}(X, y; \theta)
    \label{eq:loss}
\end{equation}
The KD loss transfers knowledge from the final to the adapted model and, as such, regularizes the training of the latter, which then improves the performance of the former.

\section{Experiments}
\begin{table*}
    \centering
    \caption{Results after learning the stream of batches from Fig. \ref{fig:subfig:s2}. All WERs (expressed in percentages) are evaluated on the final model. $\dagger$ indicates that the method's hyper-parameters were optimized on the test experiment. Best AWER result is in bold. }
    \begin{tabular}{l c c c c c c c c r}
    & & & \multicolumn{6}{c}{\textit{WER per task}}  \\
    \cmidrule(lr){4-9} 
    Model & $M$ & $(\tau, \lambda, \tau_2)$ & $\T_0$--US & $\T_1$--IRE & $\T_2$--IND & $\T_3$--AUS & $\T_4$--ENG & $\T_5$--SCO & \textbf{AWER}  \\
    \toprule
    \multicolumn{3}{l}{Initial model $\theta_0$} & 17.3 & 11.0 & 21.4 & 15.4 & 13.9 & 15.2 & 15.72 \\
    \multicolumn{3}{l}{FT} & 19.1 & 11.5 & 25.3 & 14.2 & 13.1 & 14.5 & 16.27   \\
    \midrule 
    UOE & & & 18.7 & 11.7 & 24.2 & 13.0 & 12.2 & 14.3 & 15.68 \\
    EWC$\dagger$ & & & 17.9 & 11.1 & 22.9 & 14.1 & 12.6 & 15.0 & 15.60 \\
    ER$\dagger$ & $2.0k$ &  & 17.8 & 10.8 & 21.4 & 13.9 & 12.5 & 14.2 & 15.11 \\
    O-GEM & $2.0k$ & & 19.0 & 11.5 & 25.3 & 14.0 & 13.0 & 14.3 & 16.18  \\
    \midrule 
    AOS & & $(1, 0.1, 1)$ & 17.2 & 10.7 & 21.2 & 14.3 & 12.9 & 14.4 & 15.11 \\
    AOS$\dagger$ & & $(2, 0.1, 1)$ & 17.5 & 10.7 & 21.5 & 13.7 & 12.6 & 14.3 & \textbf{15.03} \\
    \bottomrule
    \end{tabular}
    \label{tab:seq2}
\end{table*}

All experiments are done in ESPnet2 \cite{watanabe2018espnet}. For all information regarding the experiments, we refer to our Github repository \footnote{\href{https://github.com/StevenVdEeckt/online-cl-for-asr}{https://github.com/StevenVdEeckt/online-cl-for-asr}}.

\begin{figure}
\centering
\begin{subfigure}[b]{0.45\textwidth}
   \includegraphics[width=1\linewidth]{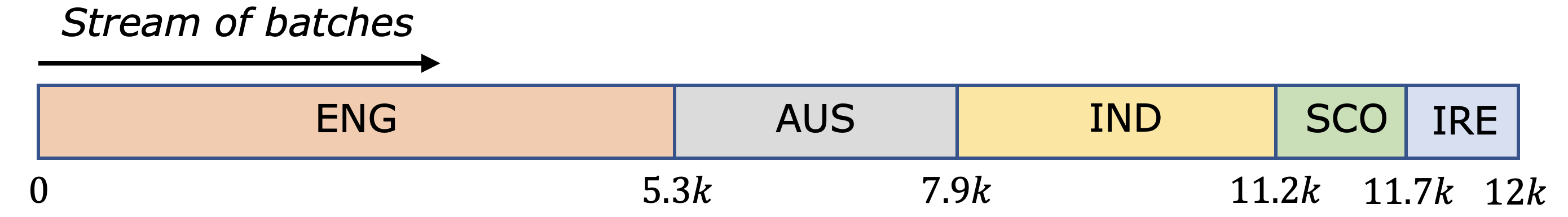}
   \caption{Sequence 1}
   \label{fig:subfig:s1} 
\end{subfigure}
\begin{subfigure}[b]{0.45\textwidth}
   \includegraphics[width=1\linewidth]{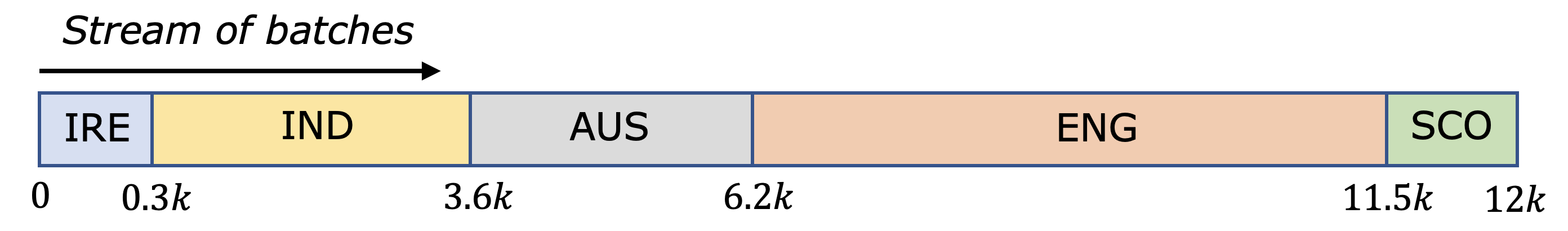}
   \caption{Sequence 2}
   \label{fig:subfig:s2}
\end{subfigure}
\caption{The stream of batches as presented to the model. The axis shows the batch number (average batch size is 22). The batches are sorted by speaker and split into five accents/tasks, as shown. The task boundaries are not known by the model.}
\label{fig:datastream}
\end{figure}

\textbf{Data.} We consider English data of Common Voice (CV) \cite{commonvoice}, split into six accents: United States (US), England (ENG), Australia (AUS), India (IND), Scotland (SCO), Ireland (IRE). The initial model $\theta_0$ is trained on an initial task $\T_0$; the batches of the remaining five tasks are sorted by task and by speaker and as such presented to the model. This makes the experiment more challenging, since the model is susceptible to forgetting both across tasks (accents) and within tasks (speakers).   We consider two sequences of the tasks. Both take US as the initial task $\T_0$, since US is by far the largest task ($350k$ utterances). We consider this to be the most realistic scenario in practice. For the five remaining tasks ($262k$ utterances), we consider the two sequences as shown in Fig. \ref{fig:datastream}: ENG$\rightarrow$AUS$\rightarrow$IND$\rightarrow$SCO$\rightarrow$IRE and IRE$\rightarrow$IND$\rightarrow$AUS$\rightarrow$ENG$\rightarrow$SCO. 

\textbf{Model.} The model ($47M$ parameters) consists of 12 Conformer \cite{conformers} encoders and 6 Transformer \cite{transformers} decoders of dimension 2048, with 4 attention heads with dimension 256. The output are $C=5000$ word pieces generated by Sentence Piece \cite{sentencepiece} on $\T_0$. The weight of the CTC loss is $c=0.3$. The model is trained on initial task $\T_0$ for 80 epochs. Afterwards, it learns the stream of batches $\B_i$ ($i>0$), seeing each batch only once. The batch size is 32, but since each batch only contains one speaker, in practice the average batch size is $22$. The model is updated with the SGD optimizer with a learning rate of $0.01$.

\textbf{Baselines.} We consider the following baselines:
\begin{itemize}
    \item \textit{Fine-Tuning (FT)}: adaptation without regularization. FT is considered the worst case baseline and will suffer from CF.
    \item \textit{Experience Replay (ER)} \cite{er}: trains jointly on the new batch and a batch sampled from memory. We consider the implementation of \cite{eeckt2021continual} with regularization weight. 
    \item \textit{Online Gradient Episodic Memory (O-GEM)} \cite{ogem}: online implementation of GEM \cite{gem}, which updates the gradient before the SGD update to prevent interference with previous tasks. We sample randomly from the memory. 
    \item \textit{Update Only Encoders (UOE)} \cite{updating_only}: proposes to only update the encoders (without layer normalization) to overcome CF.
    \item \textit{Elastic Weight Consolidation (EWC)} \cite{ewc}: computes for all parameters 'importance weights', used in a weighted L2 regularization loss. We consider the online version of \cite{pandc}.
\end{itemize}
For the rehearsal-based methods (ER and O-GEM), we consider reservoir sampling \cite{reservoir_sampling} to fill the memory of size $M=2k$. 

\textbf{Metrics.} We report the WER per task, as well as average WER (AWER), averaged over all seen tasks (accents).

\textbf{Hyper-parameters.} For the hyper-parameters, we run hyper-parameter searches on a 'test experiment', by adapting the model trained on the initial task $\T_0=\text{US}$ to some small accents from CV not present in one of the six tasks. These accents account for only $13k$ utterances, so approximately $5\%$ of the 'real' experiment. Next, we consider the AWER between the validation sets of US and the new task to find the optimal hyper-parameter value. Since the test experiment contains only a small number of utterances, we put additional focus on 'not forgetting' by giving the WER on US in the computation of AWER a higher weight than the WER on the new task (2 vs. 1). 

\section{Results}
Tables \ref{tab:seq1} and \ref{tab:seq2} show the results of the experiments for the two sequences. The performance of FT and its forgetting illustrate that CF remains a serious issue, even if the new domains (tasks) are not so dissimilar to the original one (i.e. different accents). In addition, from Tab. \ref{tab:seq1}, we observe the following:
\begin{itemize}
    \item Our method, AOS, outperforms all baselines, even with the default (i.e. non-optimized) values for its hyper-parameters. These baselines include ER and O-GEM, which have access to a memory of $M=2k$ utterances. This is a relatively large memory, given that the $262k$ utterances from the stream of batches are only seen once. If a smaller memory has to be used (e.g. $M=0.5k$), the performance of ER deteriorates and the gap with our (rehearsal-free) method becomes larger. UOE, which freezes the decoder and norm layers of the initial model $\theta_0$, is the second best baseline. It is also rehearsal-free and works well, yet is still outperformed by our method.
    \item In particular the performance of the methods on $\T_0$--US is interesting to compare, since this is the initial task the model was trained on. We can see that none of the baselines come close to the zero forgetting that our method achieves (by comparing it to \textit{Initial model $\theta_0$}), and the gap between our method and the best rehearsal-free baseline, UOE, in this regard becomes large, since UOE suffers from CF. Our method with the default setting even achieves positive backward transfer, i.e. by learning new tasks it improves on old ones. 
    \item Even in the default setting, AOS outperforms all baselines. Nevertheless, when optimizing $(\tau, \lambda, \tau_2)$ on the test experiment, we obtain even better performance, though the difference is small. With the optimal hyper-parameters, the encoder is averaged more progressively than the decoder ($\tau > \tau_2$). As a result, there is no positive backward transfer on $\T_0$--US, however, on the new tasks, the model performs better thanks to increased plasticity ($\tau>1$).  
\end{itemize}
Similar observations apply to the second sequence (Tab. \ref{tab:seq2}):
\begin{itemize}
    \item Our method outperforms all baselines, with the exception of ER for the default setting of our method; they achieve the same performance. With the optimal hyper-parameters, AOS outperforms all baselines, including ER. Note also that here UOE is much less effective than in Tab. \ref{tab:seq1}. Consequently, the gap between our method and the rehearsal-free baselines (UOE and EWC) is wide; only our method is able to compete with the rehearsal-based baselines. 
    \item Our method achieves the best performance on the initial task $\T_0$--US. Again, the default setting of our method achieves positive backward transfer, while the optimal setting achieves slight forgetting, though still better than the other methods.
    \item The optimal setting again outperforms the default setting, though the difference is small and it comes at the cost of some forgetting for the initial task. However, increased plasticity ($\tau>1$) enables the model to better learn the new tasks for the optimal setting of our method. 
\end{itemize}
Overall, it can be seen that our method is highly effective, surpassing, even in its default setting, all baselines including those with large memory and this in both sequences. In addition, our method achieves zero forgetting unlike the baselines. 

\section{Conclusions}
We propose a very effective and simple method for online CL for end-to-end ASR that involves two models: an adapted model and final model. The adapted model is trained on new batches with regularization, and its parameters are then averaged with those of the final model to update the final model. The weight of the averaging is determined by the number and length of utterances in the new batch compared to those previously seen. The averaging transfers knowledge from the adapted model to the final model, allowing it to learn new tasks without forgetting old ones. This is illustrated by the experiments,  in which our method, even in its default setting, outperforms all baselines, including those with large memory. Our method is rehearsal-free, making it simpler and more efficient than other approaches.

By overcoming the need for task boundaries and/or a memory, our method takes a step towards a more general CL method for ASR that can work in many different scenarios (when storing utterances is not allowed and/or task boundaries unknown). In future work, we aim to further extend our method in two ways, by allowing it to: (1) learn new batches in an unsupervised way; (2) introduce new word pieces to the vocabulary if needed. While this paper is an important next step towards general CL for ASR, these two objectives will be our focus in the future to achieve a truly general CL method for ASR.

\section{Acknowledgments}
Research supported by Research Foundation Flanders (FWO) under grant S004923N of the SBO programme.

\bibliographystyle{IEEEtran}
\bibliography{mybib}

\end{document}